\documentclass[aps,twocolumn,floatfix]{revtex4}%
\usepackage{amsfonts}
\usepackage{amsmath}
\usepackage{amssymb}
\usepackage{graphicx}
\usepackage{graphicx}%
\begin{document}
\title{Space-time analogy and its application to design schemes borrowed from Fourier
optics for processing ultrafast optical signals}
\author{R. N. Shakhmuratov}
\affiliation{Zavoisky Physical-Technical Institute, FRC Kazan Scientific Center of RAS,
Kazan 420029, Russia}
\date{{ \today}}

\begin{abstract}
Square-wave pulse generation with a variable duty ratio can be realized with
the help of ideas of Talbot array illuminators formulated for binary phase
gratings. A binary temporal phase modulation of CW laser field propagating
through a group-delay-dispersion circuit of the fractional Talbot length $P/Q$
results in a well defined sequence of square-wave-form pulses. When $P=1$ a
duty ratio of the pulses $D$ is $1/2$ for $Q=4$ and $1/3$ for $Q=3$ and 6.
Maximum intensity of the pulses doubles and triples compared to the CW
intensity for $D=1/2$ and $1/3$, respectively. These pulses can be used for
return-to-zero laser field modulation in optical fiber communication. For $D=1/3$
extra features between the pulses are found originating from a finite rise and
drop time of phase in a binary phase modulation. Similar effect as a benefit of
the time-space analogy is predicted for binary phase gratings and interpreted as
gleams produced by imperfect edges of the components of the rectangular phase
gratings.
\end{abstract}

\maketitle

\section{Introduction}

The space-time analogy is a useful concept that gives ideas to design new
schemes for optical signal processing in time domain borrowed from Fourier
Optics and developing in a new domain known as Temporal Optics, see review
\cite{Torres2011}. For example, a periodic phase modulated continuous wave
(CW) light, which is transmitted through a group-delay-dispersion (GDD)
circuit, has a spatial analogue corresponding to the wave field after
Fresnel diffraction by one-dimensional periodic objects in space. Flat-top
pulse generation with a duty ratio of 50\% is one of the examples of the
space-time analogy application at a quarter of the Talbot condition. These
pulses are obtained by sinusoidal \cite{Berger2004,Komukai,Torres2006}
and binary \cite{Torres2006,Fernandez,Finot} electrooptic phase modulations
of a CW laser field after transmission of the field through an appropriate
GDD circuit. This method can be used in ultrahigh-speed optical fiber
communications for energy preserving modulation of the laser field intensity
with high extinction ratio.

Mode-locked lasers are widespread sources of ultrashort optical pulses.
However, the generated pulses suffer from instability problems resulting
in jitters of the peak-power and duration. It is also difficult to control
the pulse width, shape, repetition rate, and position in a time slot for
fine tuning and synchronization with other electrical signals in the
integrated circuit. These problems can be solved by use of electrooptic
phase modulators, which allow generation of ultrashort pulses from the
continuous wave (CW) light emitted from a narrowband stabilized laser.
Electrooptic modulation transforms CW light to phase-modulated light with
ultrawide optical sidebands on the order of terahertz. Then, after
transmission through a group-delay-dispersion (GDD) circuit this chirped
CW light is compressed into trains of ultrashort pulses in the sub-picosecond
range \cite{Murata}. Meanwhile, the sinusoidally phase-modulated light with
a large phase-modulation index is not fully compressed into pulses
\cite{Otsuji}. Half of the field energy is left in a dc floor level between
pulses \cite{Torres2006}. Therefore, conditions for return-to-zero laser
field modulation without losses need to be analyzed.

In space domain periodic phase gratings are capable to convert a uniformly
wide beam losslessly into an array of periodic spots of concentrated light
\cite{Lohmann1988,Lohmann1990,Leger,Arrizon1992,Arrizon1993,Arrizon1994}. This
is so called an array illuminator (AIL), which provides illumination for
microcomponents such as optical logic gates or bistable elements in a $2-D$
discrete parallel processor \cite{Streibl1988}. AIL has aplications in
multiple imaging \cite{Dammann1971} and optical interconnectors
\cite{LohmanLoh}. In AIL a uniformly wide beam is transformed by a laterally-periodic
phase grating under paraxial approximation into the Fresnel diffraction pattern
at a distance $z$ if $z=(P/Q)Z_{T}$, where $P$ and $Q$ are coprime integers and
$Z_{T}$ is the Talbot length. This is known as the spatial fractional Talbot effect.
With a binary phase grating as the input one can create arrays of many periodic
light spots with varying spot shapes at the fractional imaging distances
\cite{Lohmann1988,Lohmann1990,Leger,Arrizon1992,Arrizon1993,Arrizon1994}. In
this paper a binary phase modulation of CW laser field and successive phase
treatment by a GDD circuit is considered. Here, a phase modulation substitutes
the phase grating, while the GDD circuit simulates Fresnel diffraction. The
fractional Talbot effect, discussed in this paper, gives a wide variety of
possibilities to generate and control ultrashort optical pulses by integrated
optical components.

\section{Binary phase-modulated CW field}

We consider the CW radiation field $E_{M}(t)=E_{0}\exp[-i\omega_{r}t+ikz+i\varphi
(t)]$ with a periodic binary phase modulation, which is%
\begin{equation}
\varphi(t)=\sum_{n=-\infty}^{+\infty}\phi(t-nT), \label{Eq1}%
\end{equation}
where%
\begin{equation}
\phi(t)=\Delta\left[  \theta\left(  t-T_{-}\right)  -\theta\left(
t-T_{+}\right)  \right]  . \label{Eq2}%
\end{equation}
Here, $E_{0}$ is the constant field amplitude, $\omega_{r}$ and $k$ are the
carrier frequency and the wave number of the field, respectively, $\theta(t)$
is the Heaviside step function, $\Delta$ is the maximum phase shift induced by
an electrooptic modulator, $T_{\pm}=(T\pm T_{p})/2$, $T$ is a modulation
period, and $T_{p}$ is a duration of the phase shift. Below we will consider
three cases when $T_{p}/T=1/2$, $1/3$, and $2/3$, see Fig. 1(a). These cases
have previously been explored in the field of Fourier Optics to design binary
phase gratings known as array illuminators that produce a two-dimensional
array of high contrast bright spots in the spatial irradiance distribution at
the fractional Talbot length without losses \cite{Arrizon1992,Arrizon1993}. In
this paper the findings in Fourier Optics (spatial domain) will be applied
to generate short pulses in Temporal Optics (time domain) producing the
irradiance concentration in well defined time slots without losses.
\begin{figure}[ptb]
\resizebox{0.3\textwidth}{!}{\includegraphics{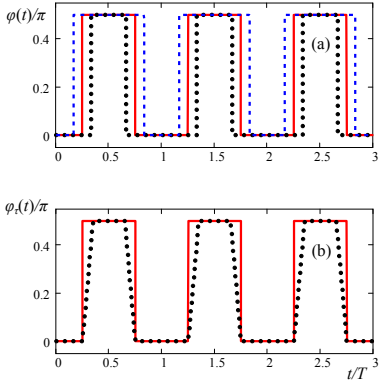}}
\caption{(a) Binary phase modulation $\varphi(t)$. $T_{p}/T$ is $1/2$ (solid
red line), $1/3$ (dotted black line), and $2/3$ (dashed blue line). (b) Comparison
of $\varphi(t)$ (red solid line) with $\varphi_{\tau}(t)$ (black dotted line)
for $T_{p}/T=1/2$ and $\tau/T=0.1$. Phase shift $\Delta$ is $\pi/2$ in both plots,
(a) and (b).}%
\label{fig_sim}%
\end{figure}

Fourier transform%
\begin{equation}
\frac{1}{T}\int_{0}^{T}e^{i\varphi(t)+i2\pi nt/T}dt=C_{n} \label{Eq3}%
\end{equation}
gives the Fourier content of the field $E_{M}(t)$, which can be expressed as
$E_{M}(t)=E_{0}e(t)e^{-i\omega_{r}t+ikz}$, where
\begin{equation}
e(t)=\sum_{n=-\infty}^{n=\infty}C_{n}e^{-i2\pi nt/T}, \label{Eq4}%
\end{equation}%
\begin{equation}
C_{n}=(-1)^{n}a\frac{T_{p}}{T}\text{sinc}\left(  \pi n\frac{T_{p}}{T}\right)
\label{Eq5}%
\end{equation}
for $n\neq0$, $a=\exp(i\Delta)-1$, sinc$\left(  y\right)  =\sin(y)/y$, and
$C_{0}=1+aT_{p}/T$ for $n=0$. Equation (\ref{Eq4}) can be simplified as%
\begin{equation}
e(t)=C_{0}+2\sum_{n=1}^{n=\infty}C_{n}\cos\left(  2\pi n\frac{t}{T}\right)  .
\label{Eq6}%
\end{equation}
According to this equation the field intensity $I_{M}(t)=\left\vert
E_{M}(t)\right\vert ^{2}$ drops to the value, which is smaller than
$I_{CW}=E_{0}^{2}$ at specific moments in time. For example, for $T_{p}/T=1/2$
we have $C_{n}=0$ for all even $n\neq0$. Therefore, at the moments of time
$t=(2m+1)T/4$, where $m$ is an integer, all the terms in the sum in Eq. (6)
are zero since $\cos\left(  2\pi n\frac{t}{T}\right)  =0$ for odd $n$. At
these moments the drop level is $I_{CW}\cos^{2}(\Delta/2)$. For example, for 
the case $\Delta=2\pi/3$ the field intensity drops to the level $I_{CW}/4$ at 
the indicated times.

This is in conflict with the definition of the phase modulated field
$E_{M}(t)=E_{0}\exp[-i\omega_{r}t+ikz+i\varphi(t)]$, which gives constant
intensity $\left\vert E_{M}(t)\right\vert ^{2}=E_{0}^{2}$ for any $t$ and any
phase modulation function $\varphi(t)$. To avoid this inconsistency we
introduce linear rise and drop of the phase substituting stepwise phase
changes in binary phase modulation. In this case the function $\phi(t)$ in Eq.
(\ref{Eq1}) is replaced by\setlength{\arraycolsep}{0.0em}
\begin{align}
\phi_{\tau}(t)  &  {}={}\Delta\left\{  \frac{t-T_{-}}{\tau}\theta
(t-T_{-})-\frac{T_{+}-t}{\tau}\theta(t-T_{+})\right. \nonumber\\
&  {+}\left(  1-\frac{t-T_{-}}{\tau}\right)  \theta(t-T_{-}-\tau)\nonumber\\
&  {-}\left.  \left(  1+\frac{t-T_{+}}{\tau}\right)  \theta(t-T_{+}%
+\tau)\right\}  , \label{Eq7}%
\end{align}
where $\tau$ is a short time interval when the phase rises or drops, see Fig.
1(b). For such a phase evolution the coefficients in the Fourier series
expansion (4) are modified for $n\neq0$ as%
\begin{align}
C_{n}  &  =\frac{(-1)^{n}}{d_{n}}\left[  e^{i\Delta}\sin b_{n}-\sin
c_{n}\right. \nonumber\\
&  {-}\left.  ig_{n}\left(  e^{i\Delta}\cos b_{n}-\cos c_{n}\right)  \right]
, \label{Eq8}%
\end{align}
and for $n=0$ as%
\begin{equation}
C_{0}=1+a\frac{T_{p}}{T}-\frac{2\tau}{T}\left[  1+a\left(  1+\frac{i}{\Delta
}\right)  \right]  , \label{Eq9}%
\end{equation}
where $b_{n}=\pi n(T_{p}-2\tau)/T$, $c_{n}=\pi nT_{p}/T$, $d_{n}=\pi n\left[
1-\left(  \frac{2\pi n\tau}{\Delta T}\right)  ^{2}\right]  $, and $g_{n}=2\pi
n\frac{\tau}{\Delta T}$. Then, the field amplitude $E_{M}(t)$ does not
experience a drop at $t=(2m+1)T/4$. To avoid time dependence of the intensity
$I_{M}(t)$, one has to take at lest $2T/\tau$ terms in the sum (\ref{Eq4}).
For example, for $T/\tau=10$, it is necessary to sum 21 terms in the interval
$(-n,n)$ where $n=10$.

In numerical simulations we should avoid fulfilment of the condition
$n^{2}=(\Delta T/2\pi\tau)^{2}$ when $d_{n}=0$ making one of the coefficients
$C_{n}$ in Eqs. (\ref{Eq4}),(\ref{Eq8}) infinite. Meanwhile, this singularity
is false since according to the definition (\ref{Eq3}) all the coefficients
$C_{n}$ are finite.

\section{Space-time analogy}

In this section we show that spatial diffraction of light from periodic
structure is analogous to the propagation of the phase modulated light through
the group-delay-dispersion circuit. The results obtained previously for the
binary phase grating are applied for generating rectangular pulses with
variable duty ratio from the phase modulated CW field.

It is well known that a uniform plane wave can be converted by binary phase
grating into many concentrated light spots in a controlable way with almost no
energy loss
\cite{Lohmann1988,Lohmann1990,Leger,Arrizon1992,Arrizon1993,Arrizon1994}. If
the grating is located at $z=0$, the plane wave $E_{S}(x,z)$ propagating along
$z$-direction is transformed immediately behind the grating for the 1-D case
as
\begin{equation}
E_{S}(x,0)/E_{0}=1+\sum_{n=-\infty}^{n=\infty}A_{n}e^{-i2\pi nx/D},
\label{Eq10}%
\end{equation}
where $A_{n}=[\exp(i\Psi)-1](W/D)\text{sinc}(\pi nW/D)$ is the $n$th Fourier
coefficient representing the $n$th complex amplitude of the grating angular
spectrum, $D$ is a grating period along $x$-direction, which is perpendicular
to the propagation direction $z$, $\Psi$ is a phase step, and $W$ is a 
width of the phase step in the binary phase grating. Mathematically $E_{S}(x,0)$ 
is fully equivalent to the periodically phase modulated field\ $E_{M}(t)$, Eq. 
(\ref{Eq4}). This can be shown by substituting $x\rightarrow t$, 
$\Psi\rightarrow\Delta$, $D\rightarrow T$, $W\rightarrow T_{p}$. The only 
difference is in $(-1)^{n}$ present in the coefficients $C_{n}$ and originating 
form the definition of a spatial grating center. This difference is easily 
removed by the center shift on a half of the grating period.

The field amplitude at a plane $z\geq0$ is transformed due to Fresnel
diffraction under the paraxial approximation as%
\begin{equation}
E_{S}(x,z)/E_{0}=1+\sum_{n=-\infty}^{n=\infty}A_{n}e^{-i2\pi n\frac{x}%
{D}-i2\pi n^{2}\frac{z}{Z_{T}}}, \label{Eq11}%
\end{equation}
where $Z_{T}=2D^{2}/\lambda$ is the Talbot length\ and $\lambda$ is the
wavelength of the field
\cite{Lohmann1990,Leger,Arrizon1992,Arrizon1993,Arrizon1994}. The
Fresnel field of the grating in the plane located at the fractional Talbot
length $Z_{T}P/Q$, where $P$ and $Q$ are coprime integers, is analyzed in
Refs. \cite{Lohmann1988,Lohmann1990,Leger,Arrizon1992,Arrizon1993,Arrizon1994}%
. To find conditions producing AIL with binary distribution of the field
$E_{S}(x,z)$, it was expressed in Ref. \cite{Arrizon1993} as binary, i.e.,
$E_{S}(x,z)=hE_{0}\sum_{n=-\infty}^{n=\infty}c_{n}\exp(-i2\pi nx/D)$, where
$h$ is a complex quantity, $c_{n}=(W^{\prime}/D)$sinc$(\pi nW^{\prime}/D)$,
$D$ is the period of the light spots, and $W^{\prime}$ is the width of the
individual light spot not necessarily equal to $W$. Analysis in Ref.
\cite{Arrizon1993} showed that binary field distribution is realized for the
following values of the parameters (i) $W/D=W^{\prime}/D=1/2$, $\Psi=\pi/2$,
$z=Z_{T}/4$, (ii) $W/D=W^{\prime}/D=1/3$, $\Psi=2\pi/3$, $z=Z_{T}/3$, and
(iii) $W/D=2/3$, $W^{\prime}/D=1/3$, $\Psi=2\pi/3$, $z=Z_{T}/6$.

Below we apply these findings in time domain and develop a method of
generating rectangular pulses with variable duty ratio. We derive simple
analytical expressions with the help of different approach developed to
calculate the defocused pattern for an arbitrary periodic grating in space,
Ref. \cite{Guigay}. We also show that imperfect edges of the binary
(rectangular) phase grating result in appearance of gleams, which are unwanted
narrow light spots between designed spots in the binary light pattern in space
and time domains.

The phase modulated field, $E_{M}(t)$, which propagates inside the GDD
circuit, is transformed at distance $z$ as, see for example Ref.
\cite{Torres2011},
\begin{equation}
E_{f}(t)=E(t)\sum_{n=-\infty}^{n=\infty}C_{n}e^{-i2\pi n\frac{t-\Phi_{1}}%
{T}+i2\pi n^{2}\frac{\Phi_{2}}{\Phi_{2T}}}, \label{Eq12}%
\end{equation}
where $E(t)=E_{0}e^{-i\omega_{r}t+ikz}$, $\Phi_{1}$ is the group delay due to
the reduced group velocity, $\Phi_{2}=\beta_{2}z$ is the GDD coefficient,
$\beta_{2}$ is the second-order dispersion coefficient, $z$ is the propagation
distance, $\Phi_{2T}=T^{2}/\pi$ is the Talbot dispersion, and $f=\left\vert
\Phi_{2}\right\vert /\Phi_{2T}$. Omitting for simplicity $\Phi_{1}$, one can
find that Eqs. (\ref{Eq11}) and (\ref{Eq12}) are equivalent if $\Phi_{2T}$ is
taken as the Talbot length and $\Phi_{2}$ is a propagation distance, both are
normalized. Below we consider the normalized fractional Talbot length
$\left\vert \Phi_{2}\right\vert /\Phi_{2T}=P/Q$ with $P=1$.

According to Refs. \cite{Guigay,Shakhmuratov} for $P=1$, integer $Q$, and
negative $\beta_{2}$ the field $E_{f}(t)$ is reduced to%
\begin{equation}
E_{\frac{1}{Q}}(t)=E(t)\sum_{m=1}^{Q}e^{i\left[  \varphi\left(  t+\frac{mT}%
{Q}\right)  -\pi/4+\frac{\pi m^{2}}{2Q}\right]  }\frac{\left[  1+i^{Q}%
(-1)^{m}\right]  }{\sqrt{2Q}}. \label{Eq13}%
\end{equation}

(i) In time domain this case is realized for the following values of the
parameters: $T_{p}/T=1/2$, $\Delta=\pi/2$, and $P/Q=1/4$, known as the
one-quarter Talbot condition. For $f=1/4$ the field $E_{f}(t)$, Eq.
(\ref{Eq12}), is simplified as follows (see Refs. \cite{Guigay,Shakhmuratov})%
\begin{equation}
E_{\frac{1}{4}}(t)=E(t)\frac{e^{-i\frac{\pi}{4}+i\varphi\left(  t\right)
}+e^{i\frac{\pi}{4}+i\varphi\left(  t+T/2\right)  }}{\sqrt{2}}. \label{Eq14}%
\end{equation}
Then, it is easy to calculate very simple expression for the field intensity,
$I_{f}(t)=\left\vert E_{f}(t)\right\vert ^{2}$, which is%
\begin{equation}
I_{\frac{1}{4}}(t)=I_{CW}\left[  1+\sin\psi_{0}(t)\right]  , \label{Eq15}%
\end{equation}
where $\psi_{0}(t)=\varphi(t)-\varphi(t+T/2)$. This expression was also
derived by different method in Refs. \cite{Arrizon1992,Torres2006}.

For binary phase modulation, shown in Fig. 1(b), the phase $\psi_{0}(t)$ jumps
stepwise between two values, $\Delta$ and $-\Delta$. If $\Delta=\pi/2$, then
according to Eq. (\ref{Eq15}), the field intensity varies stepwise between
$2I_{CW}$ and zero, see Fig. 2a, demonstrating flat-top pulses with a $50\%$
duty ratio ($D=1/2$) and complete extinction between pulses. The energy of the
phase modulated field is distributed between pulses with intensity $2I_{CW}$
and dark windows with zero intensity not violating the law of conservation of
energy. Pulses are produced due to constructive interference of two replicas
of the field, which are relatively shifted to each other in phase and time,
see Eq. (\ref{Eq14}). Dark windows appear due to destructive interference of
these replicas. Experimental implementation of this case is reported in Ref.
\cite{Fernandez}. These pulses with $50\%$ duty ratio are proposed for
return-to-zero amplitude modulation in optical fiber communication.

One can obtain multilevel temporal pattern if $T_{p}\neq T/2$. Example of the
case when $T_{p}=T/5$ is shown in Fig. 3(a). Rectangular shape pulses of
duration $T_{p}$ and intensity $2I_{CW}$ sitting on another rectangular pulse
of duration $T-T_{p}$ are formed. Intensity of the shoulders of the wide
rectangular pulses is $I_{CW}$. The pulses are separated by dark windows with
zero intensity and duration $T_{p}$.
\begin{figure}[ptb]
\resizebox{0.3\textwidth}{!}{\includegraphics{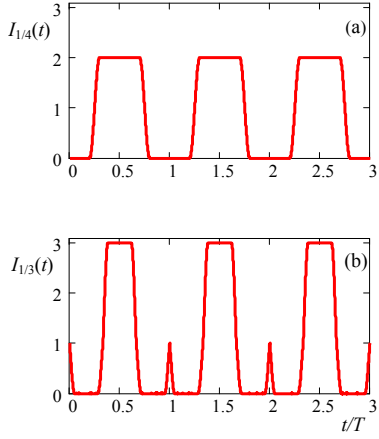}}
\caption{Time evolution of the field intensity after propagating the GDD circuit
with (a) $f=1/4$ and (b) $f=1/3$. The field is binary phase modulated with (a)
$T_{p}=T/2$, $\Delta=\pi/2$ and (b) $T_{p}=T/3$, $\Delta=2\pi/3$. Rise and
drop time of the phase is $\tau=0.05T$ in both plots.}%
\label{fig:2}%
\end{figure}

(ii) This case is realized for $T_{p}/T=1/3$, $\Delta=2\pi/3$, and $f=1/3$.
The last condition is known as the one-third Talbot condition, which gives for
Eq. (\ref{Eq13}) the following expression \cite{Shakhmuratov}%
\begin{equation}
E_{\frac{1}{3}}(t)=E(t)\frac{e^{-i\frac{\pi}{2}+i\varphi\left(  t\right)
}+e^{i\frac{\pi}{6}+i\varphi\left(  t+\frac{T}{3}\right)  }+e^{i\frac{\pi}%
{6}+i\varphi\left(  t-\frac{T}{3}\right)  }}{\sqrt{3}}. \label{Eq16}%
\end{equation}
Interference of three phase and time shifted replicas of the field in Eq.
(\ref{Eq16}) gives the following expression for the field intensity%
\begin{equation}
I_{\frac{1}{3}}(t)=\frac{I_{CW}}{3}\left[  1+8\cos\psi_{1}(t)\cos\psi
_{2}(t)\cos\psi_{3}(t)\right]  , \label{Eq17}%
\end{equation}
where $\psi_{1}(t)=[\varphi\left(  t+T/3\right)  -\varphi\left(  t-T/3\right)
]/2$, $\psi_{2}(t)=[\varphi\left(  t\right)  -\varphi\left(  t+T/3\right)
-2\pi/3]/2$, and $\psi_{3}(t)=[\varphi\left(  t\right)  -\varphi\left(
t-T/3\right)  -2\pi/3]/2$. Within time windows $(1/3+n)T<t<(2/3+n)T$ when
phase $\varphi\left(  t\right)  $ takes value $2\pi/3$ for $\tau\rightarrow0$
and an integer $n$, we have $\psi_{1}(t)=\psi_{2}(t)=\psi_{3}(t)=0$, which
gives $I_{\frac{1}{3}}(t)=3I_{CW}$. Then, in time windows $(2/3+n)T<t<(1+n)T$
when phase $\varphi\left(  t\right)  $ is zero, we have $\psi_{1}(t)=\psi
_{2}(t)=-\pi/3$ and $\psi_{3}(t)=-2\pi/3$, which gives $I_{\frac{1}{3}}(t)=0$.
In the next time window $(1+n)T<t<(4/3+n)T$ we have $\psi_{1}(t)=\pi/3$,
$\psi_{2}(t)=-2\pi/3$, and $\psi_{3}(t)=-\pi/3$, which gives again
$I_{\frac{1}{3}}(t)=0$. Thus, binary phase modulated field with $T_{p}/T=1/3$
and $\Delta=2\pi/3$ is transformed by the GDD circuit with $f=1/3$ to the sequence
of flat-top pulses with intensity $3I_{CW}$, $33\%$ duty ratio ($D=1/3$), and
complete extinction between pulses, see Fig. 2b.
\begin{figure}[ptb]
\resizebox{0.25\textwidth}{!}{\includegraphics{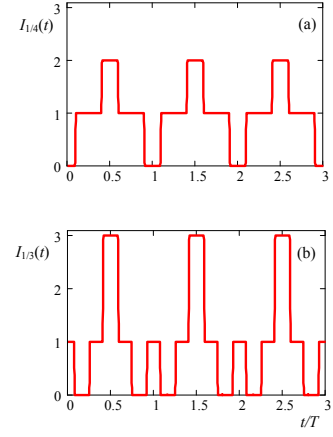}}. \caption{Time
evolution of the field intensity after propagating the GDD circuit with (a)
$f=1/4$ and (b) $f=1/3$. The field is binary phase modulated with (a)
$T_{p}=T/5$, $\Delta=\pi/2$ and (b) $T_{p}=T/2$, $\Delta=2\pi/3$. Rise and
drop time of the phase is $\tau=0.01T$ in both plots.}%
\label{fig:3}%
\end{figure}

If the rise and drop time $\tau$ of phase $\varphi(t)$ is finite, some extra
features appear in time dependence of the field intensity $I_{\frac{1}{3}}%
(t)$. They originate from the changes of phases $\psi_{1}(t)$, $\psi_{2}(t)$,
and $\psi_{3}(t)$, which take place in the vicinity of $t=nT$, i.e., $\psi
_{1}(t)$ rises from $-\pi/3$ to $\pi/3$, $\psi_{2}(t)$ drops from $-\pi/3$ to
$-2\pi/3$, and $\psi_{3}(t)$ rises from $-2\pi/3$ to $-\pi/3$.

For the linear time dependencies of the field phase $\varphi_{\tau}\left(
t\right)  $, which substitute stepwise changes in Eqs. (\ref{Eq1}%
),(\ref{Eq2}), phases $\psi_{1}(t)$, $\psi_{2}(t)$, and $\psi_{3}(t)$ change
linearly within time intervals specified below. Phase $\psi_{1}(t)=\Delta
(t-nT)/2\tau$ rises in a time interval $(nT-\tau,nT+\tau)$, phase $\psi
_{2}(t)=-\Delta(t-nT+\tau)/2\tau$ drops in a time interval $(nT,nT+\tau)$, and
phase $\psi_{3}(t)=-\Delta\lbrack1-(t-nT)/\tau]/2$ rises in a time interval
$(nT-\tau,nT)$. Such a change of phases $\psi_{1}(t)$, $\psi_{2}(t)$, and
$\psi_{3}(t)$ results in appearance of almost triangular pulses with maximum
intensity $I_{CW}$ at $t=nT$ and duration $2\tau$ at their foot.

Multilevel temporal pattern appears if $T_{p}\neq T/3$. Example of one of
these patterns is shown in Fig. 3(b) for $T_{p}=T/2$. Two-level sequence of
rectangular pulses with the period $T$ and intensities $3I_{CW}$ and $I_{CW}$
are generated. For $T_{p}=T/2$ duration of the pulses with intensity $3I_{CW}$
is $T/6$. These pulses are sitting on rectangular pulses with duration $T/2$.
Intensity of the shoulders of these long pulses is $I_{CW}$. There are also
short pulses with intensity $I_{CW}$ and duration $T/6$, which are located
between two-level pulses. Duration of the dark windows separating these short
pulses from the two-level pulses is $T/6$. In general, durations of the pulses
and dark windows in the multilevel temporal pattern depends on the ratio $T_{p}/T$.

(iii) This case is realized for $T_{p}/T=2/3$, $\Delta=2\pi/3$, and $f=1/6$,
i.e, for the one-sixth Talbot condition, which gives for Eq. (\ref{Eq13}) the
following expression \cite{Shakhmuratov}\setlength{\arraycolsep}{0.0em}
\begin{align}
E_{\frac{1}{6}}(t)  &  {}={}\frac{E(t)}{\sqrt{3}}\left\{  e^{i\frac{\pi}%
{2}+i\varphi\left(  t+\frac{T}{2}\right)  }\right. \nonumber\\
&  {+}\left.  e^{-i\frac{\pi}{6}}\left[  e^{i\varphi\left(  t+\frac{T}%
{6}\right)  }+e^{i\varphi\left(  t-\frac{T}{6}\right)  }\right]  \right\}  ,
\end{align}
Time dependence of the field intensity is%
\begin{equation}
I_{\frac{1}{6}}(t)=\frac{I_{CW}}{3}\left[  1+8\cos\psi_{4}(t)\cos\psi
_{5}(t)\cos\psi_{6}(t)\right]  , \label{Eq19}%
\end{equation}
where $\psi_{4}(t)=[\varphi\left(  t+T/6\right)  -\varphi\left(  t-T/6\right)
]/2$, $\psi_{2}(t)=[\varphi\left(  t+T/2\right)  -\varphi\left(  t+T/6\right)
+2\pi/3]/2$, and $\psi_{3}(t)=[\varphi\left(  t+T/2\right)  -\varphi\left(
t-T/6\right)  +2\pi/3]/2$. This binary phase modulated field with
$T_{p}/T=2/3$ and $\Delta=2\pi/3$ is transformed by GDD circuit with $f=1/6$
to the same sequence of flat-top pulses as in case (ii), which is shown in
Fig. 2b. It is interesting to notice that phases $\psi_{4}(t)$, $\psi_{5}(t)$,
and $\psi_{5}(t)$ are very close to those, which are introduced in case (ii),
i.e., $\psi_{4}(t)\approx\psi_{1}(t)$, $\psi_{5}(t)\approx-\psi_{3}(t)$, and
$\psi_{6}(t)\approx-\psi_{2}(t)$. There are small time shifts between these
functions in particular time intervals. However, they take place such that
time shifts with delay and advance compensate each other in Eq. (\ref{Eq19})
giving exactly the same time dependence as for $I_{\frac{1}{3}}(t)$.

Experimental implementation of this case ($T_{p}/T=2/3$, $\Delta=2\pi/3$, and
$f=1/6$) is reported in Ref. \cite{Fernandez} for $\tau\approx0.1T$. Bell
shaped pulses originating from finite rise-fall time of binary phase
modulation are clearly seen between flat-top pulses in the experiment. It
should be noted that parameter $q$ defined in Ref. \cite{Fernandez} is related
to $Q$ as $q=Q/2$.

In spatial domain these extra features between designed spots of binary form
can be considered as a glare produced by edges of the rectangular phase
grating. If the phase linearly grows or decreases at these edges between zero
and $\Psi$, light spots of triangular form appear between rectangular spots.
Maximum intensity of the spots equals to the intensity of the incident
radiation. The spot size equals to the sum of the edge lengths of the
individual phase grating element.

Multilevel temporal pattern also appears in the case $f=1/6$ if $T_{p}%
\neq2T/3$. For example, if $T_{p}=T/2$ this pattern is identical to that
considered in (ii).

\section{Conclusion}

Square-wave pulse generation with variable duty ratio based on the temporal
fractional Talbot effect is analyzed for binary phase modulated fields. Simple
expressions for the field intensity after propagating through GDD circuit are
presented for $P/Q$ fractional Talbot condition with $P=1$, $Q=$ $3$, $4$, and
$6$. Space-time analogy helps to find conditions when rectangular pulses with
variable duty ratio are generated. This method allows energy preserving
conversion of CW to pulse trains with repetition rates in the gigahertz range
and sub-nanosecond pulse width. It is shown that finite time of rise and drop
in binary phase modulation results in short, low intensity pulses between
pulses of designed sequence for fractional Talbot effect with $P/Q=3$ and $6$.
Similar effect takes place in space domain. Generation of the multilevel
temporal pattern of pulses with high extinction ratio is proposed.

\section{Acknoledgment}

This work was supported by FRC Kazan Scientific Center of RAS.

\end{document}